\documentclass[12pt]{article}
\usepackage{amsfonts,amsmath,amssymb,epsf}
\usepackage{amsmath,braket}

\usepackage{graphicx,hyperref,color}
\hypersetup{
    bookmarks=true,         % show bookmarks bar?
    unicode=false,          % non-Latin characters in Acrobat’s bookmarks
    pdftoolbar=true,        % show Acrobat’s toolbar?
    pdfmenubar=true,        % show Acrobat’s menu?
    linktocpage=true,       % hyperlink by page number in contents
    pdffitwindow=false,     % window fit to page when opened
    pdfstartview={FitH},    % fits the width of the page to the window
    pdfnewwindow=true,      % links in new PDF window
    colorlinks=true,       % false: boxed links; true: colored links
    linkcolor=blue,          % color of internal links (change box color with linkbordercolor)
    citecolor=blue,        % color of links to bibliography
    filecolor=blue,      % color of file links
    urlcolor=blue           % color of external links
}
\numberwithin{equation}{section}									% equation numbering by section

\def\d#1{\,{\rm d}#1}

\newcommand{\de}{\partial}
\newcommand{\be}{\begin{equation}}
\newcommand{\ba}{\begin{eqnarray}}
\newcommand{\ea}{\end{eqnarray}}
\newcommand{\ee}{\end{equation}}

\newcommand{\s}{\sqrt}
\newcommand{\vp}{\varphi}

\newcommand{\ti}{\tilde}
\newcommand{\ap}{\alpha}

\newcommand{\no}{\nonumber \\}
\newcommand{\la}{\langle}
\newcommand{\lb}{\rangle}
\newcommand{\bea}{\begin{eqnarray}}
\newcommand{\eea}{\end{eqnarray}}
\newcommand{\bes}{\begin{equation*}}
\newcommand{\beas}{\begin{eqnarray*}}
\newcommand{\eeas}{\end{eqnarray*}}
\newcommand{\bas}{\begin{array*}}
\newcommand{\eas}{\end{array*}}
\newcommand{\ees}{\end{equation*}}
\newcommand{\nn}{\nonumber}
\newcommand{\p}{\partial}
\newcommand{\ep}{\epsilon}

\newcommand{\Tr}{\textrm{Tr}}
\newcommand{\Sch}{\textrm{Sch}}
\let\a=\alpha \let\b=\beta \let\c=\chi \let\d=\delta \let\e=\epsilon  \let\g=\gamma \let\h=\eta \let\k=\kappa \let\l=\lambda \let\m=\mu \let\n=\nu
 \let\p=\phi \let\r=\rho %\let\s=\sigma 
 \let\th=\theta  \let\vp=\varphi   
          
\def\nn{\nonumber}
\def\inf{\infty}

\def\pa{\partial}

\newcommand{\JT}{\textrm{JT}}
\newcommand{\bulk}{\textrm{bulk}}
\newcommand{\bdy}{\textrm{bdy}}
\newcommand{\ct}{\textrm{ct}}
\newcommand{\onshell}{\scriptsize\mbox{on-shell}}
\newcommand{\oneloop}{\scriptsize\mbox{one-loop}}

\begin{document}

\begin{titlepage}
\thispagestyle{empty}

\vspace*{-2cm}
\begin{flushright}
YITP-21-88
\\
IPMU21-0054
\\
\end{flushright}

\bigskip

\begin{center}
\noindent{{\Large \textbf{JT Gravity Limit of Liouville CFT and Matrix Model}}}\\
\vspace{2cm}

\quad Kenta Suzuki$^a$ \ and \ Tadashi Takayanagi$^{a,b,c}$
\vspace{1cm}\\

{\it $^a$Center for Gravitational Physics,\\
Yukawa Institute for Theoretical Physics,
Kyoto University, \\
Kitashirakawa Oiwakecho, Sakyo-ku, Kyoto 606-8502, Japan}\\
\vspace{1mm}
{\it $^b$Inamori Research Institute for Science,\\
620 Suiginya-cho, Shimogyo-ku,
Kyoto 600-8411 Japan}\\
\vspace{1mm}
{\it $^{c}$Kavli Institute for the Physics and Mathematics
 of the Universe (WPI),\\
University of Tokyo, Kashiwa, Chiba 277-8582, Japan}\\

\bigskip \bigskip
\vskip 2em
\end{center}

\begin{abstract}
In this paper we study a connection between Jackiw-Teitelboim (JT) gravity on two-dimensional anti de-Sitter spaces
and a semiclassical limit of $c<1$ two-dimensional string theory.
The world-sheet theory of the latter consists of a space-like Liouville CFT coupled to a non-rational CFT defined by a time-like Liouville CFT. We show that their actions, disk partition functions and annulus amplitudes perfectly agree with each other, where the presence of boundary terms plays a crucial role. We also reproduce the boundary Schwarzian theory from the Liouville theory description.
Then, we identify a matrix model dual of our two-dimensional string theory with a specific time-dependent background in $c=1$ matrix quantum mechanics. 
Finally, we also explain the corresponding relation for the two-dimensional de-Sitter JT gravity.

\end{abstract}

\end{titlepage}

\newpage

\tableofcontents

%%%%%%%%%%%%%%%%%%%%%%%%%%%%%%%%%%%
\section{Introduction}
%%%%%%%%%%%%%%%%%%%%%%%%%%%%%%%%%%%

Jackiw-Teitelboim (JT) gravity is a two-dimensional theory of quantum gravity coupled to a real scalar field \cite{Jackiw:1984je, Teitelboim:1983ux}.
This model played a crucial role for our recent understanding of (nearly) AdS$_2$/CFT$_1$ correspondence
\cite{Almheiri:2014cka, Jensen:2016pah, Maldacena:2016upp, Engelsoy:2016xyb}
and it can be viewed as a leading low temperature universal sector of higher dimensional near-extremal black holes \cite{Ghosh:2019rcj, Sachdev:2019bjn}.
The model also brought us improved insights for black hole information paradox via replica wormholes \cite{Penington:2019kki, Almheiri:2019qdq, Balasubramanian:2020coy, Goto:2020wnk}.
JT gravity is especially remarkable when it is defined on a Euclidean negatively curved backgrounds, including a sum over higher genus topologies in the bulk.
In this set-up, it was shown that its dual is given by a double-scaled Hermitian matrix integral \cite{Saad:2019lba}.
The discovery of this duality led many authors further investigations, for example, generalization to other matrix ensembles \cite{Stanford:2019vob},
correlation functions \cite{Saad:2019pqd} \footnote{Related works also include \cite{Kimura:2020zke, Kimura:2021kzr}.},
non-perturbative effects \cite{Johnson:2019eik}, and fixing matrix eigenvalues by eigenbranes \cite{Blommaert:2019wfy},
as well as attempts of this type of averaged duality to higher dimensions (in particular AdS$_3$/CFT$_2$)
\cite{Afkhami-Jeddi:2020ezh, Maloney:2020nni, Cotler:2020ugk, Cotler:2020hgz}.

In the work of \cite{Saad:2019lba}, it was also suggested that JT gravity is a particular semiclassical limit of the old minimal string theory
\cite{Ginsparg:1993is, DiFrancesco:1993cyw, Alexandrov:2003ut, Seiberg:2004at}.
This connection between JT gravity and the minimal string was further studied in \cite{Okuyama:2019xbv, Okuyama:2020ncd, Seiberg:2010, Mertens:2020hbs, Turiaci:2020fjj}.
An important feature of this relation is that JT gravity looks like the {\it worldsheet} theory of the minimal string,
which consists of a (space-like) Liouville CFT plus a minimal CFT (besides the ghost sector).
Hence, we might call this particular semiclassical limit of the minimal string ``JT string'' \cite{Saad:2019lba}
\footnote{A similar idea for a worldline theory was developed in \cite{Casali:2021ewu}.},
but it is not completely clear how the degrees of freedom of JT gravity arise from the minimal string.
One useful way to describe a minimal CFT, which is rational, is 
the well-known Coulomb gas formalism with a BRST truncation \cite{DiFrancesco:1997nk}
\footnote{For a recent discussion on this Coulomb gas formalism, see \cite{Kapec:2020xaj}.}.
In this paper, we will initiate such investigation by replacing the minimal CFT by a time-like Liouville CFT
\cite{Gutperle:2003xf, Strominger:2003fn, Schomerus:2003vv, Karczmarek:2003pv, Takayanagi:2004yr, McElgin:2007ak, Harlow:2011ny}.
The time-like Liouville theory is constructed by adding a Liouville potential on top of a time-like linear-dilaton theory.
As opposed to the mininal CFT, we do not perform any truncation of the spectrum in the time-like Liouville CFT and thus it is a non-rational CFT.
Our worldsheet theory consisting of a time-like Liouville theory (or a time-like linear-dilaton theory) together with the usual space-like Liouville CFT.
This is properly regarded as a time-dependent background in two dimensional string theory, rather than the minimal string
\footnote{On the other hand, the target space of the minimal string was discussed in \cite{Seiberg:2003nm, Kutasov:2004fg, Maldacena:2004sn}.}
and indeed this has a dual description by considering a time-dependent background of the well-known $c=1$ matrix quantum mechanics \cite{ Klebanov:1991qa,Ginsparg:1993is,Polchinski:1994mb} as found in \cite{Takayanagi:2004yr}.
With this knowledge as a hint, we will try to investigate the target space picture of JT string.

A different connection between JT gravity and Liouville field theory was pointed out in \cite{Mandal:2017thl}.
Also a relation between Schwarzian theory and Liouville quantum mechanics was studied in \cite{Bagrets:2016cdf, Bagrets:2017pwq, Mertens:2017mtv}
and another Liouville field theory was obtained from the large $q$ limit of the SYK model \cite{Maldacena:2016hyu, Cotler:2016fpe, Das:2020kmt}.
However, the connection between these theories and our Liouville field theory discussed in this paper is not clear.

\bigskip \bigskip

The remainder of the paper is organized as follows.
In section~\ref{sec:JT}, we summarize JT gravity computation of on-shell actions
of the hyperbolic disk, as well as of the ``double-trumpet'' (i.e. two-boundary Euclidean wormhole) geometries.
This section mostly follows the discussion of \cite{Maldacena:2016upp, Saad:2019lba}, but using more appropriate coordinates for later Liuoville theory discussion.
The on-shell action of the double-trumpet geometry is constructed by two copies of ``single-trumpet'' geometries,
and the latter is obtained by a simple analytical continuation from a punctured disk geometry (i.e. hyperbolic disk with a conical defect)
discussion in \cite{Mertens:2019tcm, Maxfield:2020ale, Witten:2020wvy, Mefford:2020vde, Turiaci:2020fjj}.

In section~\ref{sec:liouville action}, we define our double Liouville theory by a space-like and a time-like Liouville theory, whose total centralcharge is 26 coincides with that of world-sheet theory of two dimensional string theory.
In section~\ref{sec:jt limit}, we demonstrate that the action of this double Liouville theory is reduced to the JT gravity action in the semiclassical limit with appropriate field redefinitions. This argument was originally presented in \cite{Seiberg:2010, Mertens:2020hbs}.
In order to match with JT gravity result with a Dirichlet boundary condition at asymptotic AdS$_2$,
we study the FZZT brane \cite{Fateev:2000ik, Teschner:2000md} boundary condition in our double Liouville theory
\footnote{FZZT branes in JT gravity have been recently discussed in \cite{Okuyama:2021eju}.}.
With this boundary condition, we compute the classical on-shell action of our double Liouville theory for the disk geometry,
by direct evaluation of the action with on-shell background solutions in section~\ref{sec:FZZT brane},
and by semiclassical saddle-point evaluation of the minisuperspace wave functions in section~\ref{sec:minisuperspace}.
In Liouville CFT literature, the on-shell action of the double-trumpet geometry is called annulus amplitude.
We compute the annulus amplitude of our double Liouville theory by using the same methods as disk geometry in section~\ref{sec:annulus amplitude}.
Finally in section~\ref{sec:Schwarzian theory}, we evaluate the boundary action of our double Liouville theory, which leads to the Schwarzian theory as in JT gravity.

In section~\ref{sec:matrix modelf}, we present a matrix model description of 
the JT gravity limit of the two-dimensional string theory whose matter sector is described by the time-like Liouville CFT.
We argue that this is given by the $c=1$ matrix model in a time-dependent background
and we determine the corresponding Fermi surface of this background describing the eigenvalue fluctuations. 
Next, in section~\ref{sec:matrix modelc}, we study the collective field theory of this background.
This shows that the JT limit is described by a two-dimensional string theory with a time-like Liouville wall.

In section~\ref{sec:ds2}, we describe the corresponding relation for the two-dimensional de-Sitter JT gravity \cite{Maldacena:2019cbz, Cotler:2019nbi}.
This is simply archived by changing the sign of the bulk cosmological constant of the Liouville CFT. 
Hence the dual matrix quantum mechanics does not change from the one explained in section~\ref{sec:matrix model} and indeed we are able to
confirm this in our matrix quantum mechanics description.

We give our conclusions in section~\ref{sec:conclusions} and the appendices contain some of computational details and supplemental discussions.

%%%%%%%%%%%%%%%%%%%%%%%%%%%%%%%%%%%%%%%%%%%%%%%%%%%%%%%%%%%%%%%%%%%%%%
\section{JT Gravity}
\label{sec:JT}
%%%%%%%%%%%%%%%%%%%%%%%%%%%%%%%%%%%%%%%%%%%%%%%%%%%%%%%%%%%%%%%%%%%%%%
In this section, we study the partition function of JT gravity on a fixed background.
In particular, we are interested in the hyperbolic disk and double-trumpet geometries.

We write the action of JT gravity on a two dimensional manifold $M$ as follows:
\ba
I_{\JT} \, = \, -2\int_{M}\s{g} \, \Phi (R+2) \, - \, 4\int_{\de M}\s{\gamma} \, \Phi K \, + \, I_{\ct} \, ,
\label{I_JT}
\ea
where $\g$ and $K$ are the induced metric and trace of the extrinsic curvature on the boundary $\de M$.
One can also add the Einstein-Hilbert action to the above action, but it is just topological and gives a constant contribution $-S_0 \c(M)$ to the on-shell action,
where $S_0$ is a coefficient of the Einstein-Hilbert action and $\c(M)$ is the Euler characteristic of the manifold $M$.
In this case, the dialtion field must satisfy $S_0 \gg \Phi$ at any position. 
However, for a fixed background geometry as we will consider in the rest of the paper, this topological suppression factor does not play any important role.
Therefore, we will omit the Einstein-Hilbert action in the following.

The counter term is added to make the on-shell action finite and it reads \cite{Harlow:2018tqv}
\ba
I_{\ct} \, = \, 4\int_{\de M}\s{\gamma}\Phi \, .
\label{cont}
\ea

%%%%%%%%%%%%%%%%%%%%%%%%%%%%%%%%%%%%%%%%%%%%%%%%%%%%%%%%%%%%%%%%%%%%%%
\subsection{Disk geometry}
%%%%%%%%%%%%%%%%%%%%%%%%%%%%%%%%%%%%%%%%%%%%%%%%%%%%%%%%%%%%%%%%%%%%%%
Let us first consider the disk topology.
The variation of the dilation field $\Phi$ in the bulk action of (\ref{I_JT}) simply sets $R=-2$, so the background solution is (Euclidean) AdS$_2$.
Still there is a non-trivial equation for the dilation field $\Phi$ coming from the variation of the metric.
The the global coordinates of the (Euclidean) AdS$_2$ solution, which we are interested in, is given by 
	\begin{align}
        ds^2 \, &= \, \frac{4}{(1-r^2)^2}(dr^2+r^2d\theta^2) \, , \nn\\[2pt]
        \Phi \, &= \, A\cdot \frac{1+r^2}{1-r^2} \, ,   
\label{adst}
	\end{align}
where $A$ is an integration constant of the solution.
As in the usual angular coordinates, $r$ is the radial coordinate $0<r<1$, where the boundary is located at $r=1$,
and $\th$ is an angular coordinate with periodicyty $2\pi$.

We regulate the UV  divergence coming from the AdS asymptotic boundary $r\to 1$ by introducing a UV cutoff $\d$ as \cite{Maldacena:2016upp}
\ba
r \, = \, 1-\delta \, , \qquad (0<\delta\ll 1) \, ,
\ea
which leads to the asymptotic behaviour of the background solutions
	\begin{align}
        ds^2 \, &\simeq \, \frac{1}{\delta^2}(dr^2+r^2d\theta^2) \, , \nn\\[2pt]
        \Phi \, &\simeq \, \frac{A}{\delta} \, .
	\end{align}
Eventually we will be interested in the $\d \to 0$ limit.
We can choose the boundary length to be $\beta$ by redefining the angular coordinate 
\ba
    u \, = \, \frac{\beta}{2\pi} \, \theta \, ,
\ea
where $u$ can be understood as the time of the boundary theory.
Then the corresponding UV cut off $\ep$ is now defined as $\ep=\frac{\beta}{2\pi} \, \delta$, such that
	\begin{align}
        ds^2 \, \simeq \, \frac{1}{\ep^2} \, du^2 \, , \qquad \
        \Phi \, \simeq \, \frac{\bar{\Phi}}{\ep} \, ,
    \label{JT-BC}
	\end{align}
where we set the renormalized value of the dilaton: $\bar{\Phi}=A\cdot \frac{\beta}{2\pi}$.

Since the background solution satisfies $R=-2$, the bulk action of (\ref{I_JT}) vanishes on-shell.
For the remaining boundary action, by substituting the background solution, the on-shell action is evaluated as 
	\begin{align}
        I^{\onshell}_{\JT} \, &= \, -4A\int d\theta \, \frac{1+r^2}{(1+r)^2} \bigg|_{r=1-\d} \no
        \, &\simeq \, -4\pi A=-8\pi^2\frac{\bar{\Phi}}{\beta} \, , 
    \label{jtac}
	\end{align}
where for the second line, we neglected order $\d$ contributions.

%%%%%%%%%%%%%%%%%%%%%%%%%%%%%%%%%%%%%%%%%%%%%%%%%%%%%%%%%%%%%%%%%%%%%%
\subsection{Euclidean wormhole geometry}
\label{JTwormhole}
%%%%%%%%%%%%%%%%%%%%%%%%%%%%%%%%%%%%%%%%%%%%%%%%%%%%%%%%%%%%%%%%%%%%%%
Next we would like to study two-boundary Euclidean wormhole geometry (i.e. ``double-trumpet'').
As shown in \cite{Saad:2019lba}, such geometry can be constructed by sewing two copies of ``single-trumpet''.
Also it is known that single-trumpet geometry is related to the punctured disk (i.e. disk with a conical singularity) by analytical continuation of its conical deficit angle \cite{Maxfield:2020ale, Witten:2020wvy}.
It seems more convenient to take this detour route via punctured disk (in particular for comparison with the later Liouville theory discussion),
where we just need to insert a bulk ``vertex operator'', instead of directly studying the single-trumpet geometry.

Effectively, this means that we need to study a bulk one-point function of the form $\big\la e^{-4\pi b \a\Phi(0)} \big\lb$,
\footnote{The minus sign compared to the Liouville vertex operator comes from the overall sign difference of the bulk action.}
where the operator is inserted at the center of disk and the defect angle is $\pi b \a$.
Now this modifies the JT bulk action as 
	\begin{align}
		I_{\JT}^{\bulk} \, = \, -2\int_{M}\s{g}(R+2)\Phi \, + \, 4\pi b \a \Phi(0) \, ,
	\end{align}
while the boundary action and counterterm do not change.
The corresponding background solution is given by
	\begin{align}
		ds^2 \, &= \, \frac{4\n^2 \, r^{2\n-2}}{(1-r^{2\n})^2} \, (dr^2+r^2d\theta^2) \, , \nn\\[2pt]
		\Phi \, &= \, A\cdot \frac{1+r^{2\n}}{1-r^{2\n}} \, ,
	\end{align}
where we defined 
	\begin{align}
		\n \, = \, 1- \frac{b \a}{2} \, .
	\label{nu}
	\end{align}

The on-shell action can be computed from the boundary action as before
	\begin{align}
		I^{\onshell}_{\JT} \, &= \, -4\n A\int d\theta \frac{1+r^{2\n}}{(1+r^\n)^2}\bigg|_{r=1-\d} \no
		&\simeq \, -4\pi \n^2 A=-8\pi^2 \n^2 \frac{\bar{\Phi}}{\beta} \, .
	\label{JT-pdisk}
	\end{align}
This indeed agrees with the tree-level punctured disk partition function \cite{Mertens:2019tcm}.
The single-trumpet partition function is obtained by 
	\begin{align}
		2\pi \n \, \to \, i \tilde{\a} \, ,
	\end{align}
where $\tilde{\a}$ is the geodesic length of the other end.

Finally the tree-level contribution to the double-trumpet partition function is given by \cite{Saad:2019lba}
	\begin{align}
		\int_0^{\inf} d\tilde{\a} \, \tilde{\a} \, e^{- 2\tilde{\a}^2 \frac{\bar{\Phi}}{\beta_1}} e^{- 2\tilde{\a}^2 \frac{\bar{\Phi}}{\beta_2}} \, = \, \frac{\b_1 \b_2}{4\bar{\Phi}(\b_1 +\b_2)} \, .
	\end{align}

%%%%%%%%%%%%%%%%%%%%%%%%%%%%%%%%%%%%%%%%%%%%%%%%%%%%%%%%%%%%%%%%%%%%%%
\subsection{One-loop partition function}
%%%%%%%%%%%%%%%%%%%%%%%%%%%%%%%%%%%%%%%%%%%%%%%%%%%%%%%%%%%%%%%%%%%%%%
Next, we would like to study one-loop contributions to the partition function.
Since the pure JT gravity is one-loop exact \cite{Stanford:2017thb, Saad:2019lba}, this completes the partition function.
To this end, we promote the bulk coordinates to functions of the ``boundary time'' $u$ as $r \to r(u)$ and $\theta \to \theta(u)$ \cite{Maldacena:2016upp}.
This is simply a reparametrization between the bulk Euclidean time near the boundary and the physical boundary time $u$.
Now, because of the boundary condition (\ref{JT-BC}), the two functions are not independent, but
	\begin{align}
        \qquad g_{uu} \, = \, \frac{1}{\e^2} \, = \, \frac{4(r'^2+r^2 \th'^2)}{(1-r^2)^2} \qquad \Rightarrow \qquad
        r(u) \, = \, 1 - \e \, \th'(u) \, + \, \mathcal{O}(\e^2) \, ,
	\label{r(u)}
	\end{align}
where the prime denotes a derivative respect to the boundary time $u$.
Using this relation, we would like to evaluate the on-shell action (\ref{jtac}) up to $\mathcal{O}(\e^0)$ order.
To do this, we need to be careful because the boundary cutoff is now located at $ r(u) = 1 - \e \th'(u)$.
Detail evaluation of the extrinsic curvature on this cutoff can be found in Appendix~A of \cite{Sachdev:2019bjn}, and this leads to 
	\begin{align}
	    K \, = \, 1 \, + \, \e^2 \, \Sch\big( \tan(\tfrac{\th}{2}) \, , u \big) \, + \, \mathcal{O}(\e^3) \, ,
	\end{align}
where the Schwarzian derivative is defined by
	\begin{align}
	    \Sch(F , x) \, = \, \frac{F'''(x)}{F'(x)} \, - \, \frac{3}{2} \left( \frac{F''(x)}{F'(x)} \right)^2 \, .
	\end{align}
Therefore, we find the Schwarzian action
	\begin{align}
		I^{\bdy}_{\JT} \, = \, - 4\bar{\Phi} \int_0^\b du \, \Sch\big( \tan(\tfrac{\th}{2}) \, , u \big) \, .
    \label{I_Sch}
	\end{align}

From this action, expanding around the saddle-point solution as $\th(u)=2\pi u/\b+\e(u)$, the quadratic action is obtained as
	\begin{align}
		I^{(2)}_{\JT} \, &= \, \frac{\bar{\Phi}\b^2}{2\pi^2} \int_0^\b du \, \left[ \e''{}^2(u) \, - \, \left( \frac{2\pi}{\b} \right)^2 \e'{}^2(u) \right] \no
		&= \, \frac{8\pi^2 \bar{\Phi}}{\b} \sum_m \, m^2(m^2-1) \, \e_m^2 \, , 
	\end{align}
where we used Fourier expansion $\e(u) = \sum_m e^{2\pi i m u/\b} \, \e_m$.
For the disk geometry, among the Fourier modes, we have to exclude the zero modes corresponding $m =0, \pm 1$ \cite{Stanford:2017thb}.
Combining with the appropriate measure factor \cite{Saad:2019lba}, now the one-loop partition function is obtained as
	\begin{align}
        Z_{\oneloop} \, = \, \left( \frac{\bar{\Phi}}{2\pi \b} \right)^{\frac{3}{2}} \, .
    \label{JTone-loop}
	\end{align}

%%%%%%%%%%%%%%%%%%%%%%%%%%%%%%%%%%%%%%%%%%%%%%%%%%%%%%%%%%%%%%%%%%%%%%
\section{Liouville CFT}
\label{sec:liouville cft}
%%%%%%%%%%%%%%%%%%%%%%%%%%%%%%%%%%%%%%%%%%%%%%%%%%%%%%%%%%%%%%%%%%%%%%
In this section, we define our double Liouville theory by a combination of a space-like and a time-like Liouville theory
and show that such theory agrees with JT gravity in the semiclassical limit.

%%%%%%%%%%%%%%%%%%%%%%%%%%%%%%%%%%%%%%%%%%%%%%%%%%%%%%%%%%%%%%%%%%%%%%
\subsection{Liouville action}
\label{sec:liouville action}
%%%%%%%%%%%%%%%%%%%%%%%%%%%%%%%%%%%%%%%%%%%%%%%%%%%%%%%%%%%%%%%%%%%%%%

We consider the CFT consists of a space-like Liouville field $\phi$ and the time-like one $\chi$, defined by the bulk action 
	\begin{align}
        I^{\bulk}_{\rm L} \, = \, \int d^2 x \sqrt{g} \, \Big[(\de_\m\phi)^2 - (\de_\m\chi)^2 + \mu e^{2b\phi} - \mu e^{-2b\chi} \Big] \, ,
    \label{LVb}
	\end{align}
on a flat space R$^2$
\ba
ds^2 \, = \, dr^2 \, + \, r^2d\theta^2 \, .
\label{angular coord}
\ea
Here we choose the renormalized bulk cosmological constant of the time-like Liouville theory as $-\m$ in order to agree with JT gravity \cite{Seiberg:2010, Mertens:2020hbs}.
The background charges are fixed as $Q=b+1/b$ for the space-like Liouville field $\phi$ and $q=1/b-b$ for the time-like field $\chi$ with a parameter $0<b \le1$.
Therefore, each theory has its central charge $c_\phi = 1+6Q^2$ and $c_\chi = 1-6q^2$,
so that the total central charge is $c=c_\phi+c_\chi=26$, consistent with the anormaly cancellation condition of the bosonic string theory.

The string coupling behaves as $g_s=e^{Q\phi+q\chi}$,
and it is useful to note that the conformal dimensions of bulk vertex operators read
\ba
h\big(e^{2\ap\phi} \big) \, = \, \ap(Q-\ap) \, , \qquad  h\big(e^{2\ap\chi} \big) \, = \, \ap(\ap-q) \, .
\ea
This shows that the bulk vertex operator of the form $e^{Q\phi+q\chi+2iP\phi\pm 2iP\chi}$ is dimension one.

In the presence of boundary we need to add the boundary term
\ba
I^{\bdy}_{\rm L} \, = \, 2\int d\theta(Q\phi+q\chi)K^{(0)}, 
\label{LVbd}
\ea
where $K^{(0)}$ is the trace of the extrinsic curvature in the flat space at $r=1$.

%%%%%%%%%%%%%%%%%%%%%%%%%%%%%%%%%%%%%%%%%%%%%%%%%%%%%%%%%%%%%%%%%%%%%%
\subsection{JT gravity as a semiclassical limit of Liouville action}
\label{sec:jt limit}
%%%%%%%%%%%%%%%%%%%%%%%%%%%%%%%%%%%%%%%%%%%%%%%%%%%%%%%%%%%%%%%%%%%%%%
In this subsection, we will explain the semiclassical limit of the double Liouville theory, which gives the JT gravity result.
(See also appendix~F of \cite{Mertens:2020hbs}.)

We introduce field redefinition of the two Liouville fields by
	\begin{align}
		\phi \, = \, \rho/b-b\Phi \, , \qquad &\chi \, = \, - \, \rho/b-b\Phi \, , \no
        \rho \, = \, \frac{b}{2}(\phi-\chi) \, , \qquad &\Phi \, = \, - \, \frac{1}{2b}(\phi+\chi) \, .
    \label{rho&Phi}
	\end{align}
This redefinition gives us 
\footnote{
In order to derive this boundary action, we {\it first} approximated $Q \approx 1/b$ and $q \approx 1/b$ in the boundary action, then used (\ref{rho&Phi}).
Instead if we use the original $Q$ and $q$ at the same time as (\ref{rho&Phi}), this leads to an additional term containing $\rho K^{(0)}$.
This term does not change the finite contributions of the following discussion, but leaves slight subtlety for the counter term.
We will comment on this contribution in appendix~\ref{app:rho K0}.
\label{foot:rho K0}
}
	\begin{align}
	    I^{\bulk}_{\rm L} \, &= \, \int d^2 x \sqrt{g} \, \Big[ -4(\pa \r) \cdot (\pa \Phi) \, - \, 2 \m e^{2\r} \sinh(2b^2 \Phi) \Big] \, , \\
	    I^{\bdy}_{\rm L} \, &= \, -4\int d\theta \, \Phi K^{(0)} \, .
	\end{align}
Now, we regard $\rho$ as the conformal factor of the JT gravity metric as 
\ba
ds_{\JT}^2 \, = \, e^{2\rho}(dr^2+r^2d\theta^2) \, .
\label{conformal gauge}
\ea 
After including this Weyl factor and writing in terms of the JT metric, we find 
	\begin{align}
	    I^{\bulk}_{\rm L} + I^{\bdy}_{\rm L} \, = \, -2 \int d^2 x \sqrt{g_{\JT}} \, \Big[ \Phi R + \m \sinh(2b^2 \Phi) \Big]
	    \, - \, 4\int d\theta\s{\g_\JT} \, \Phi K \, ,
	\end{align}
where $R$ is the Ricci scalar for the JT gravity metric.
In the semiclassical ($b\to 0$) limit with $\mu b^2=1$, we can see that the Liouville theory defined by (\ref{LVb}) and (\ref{LVbd}) gets equivalent to JT gravity.
This type of dilaton-gravity model was also studied in \cite{Kyono:2017jtc, Kyono:2017pxs, Okumura:2018xbh} as a Yang-Baxter deformation of JT gravity,
where parameter $b$ measures the deformation. It would be interesting to understand this connection, but we leave this for a future work.

The counter term (\ref{cont}) in JT gravity is expressed in terms of Liouville fields as
\ba
I_{ct} \, = \, 4\int d\theta \sqrt{\g} \left[-\frac{1}{2b}(\phi+\chi)e^{\frac{b}{2}(\phi-\chi)}\right] \, .
\label{I_ct1}
\ea
In the semiclassical $b\to 0$ limit, we can employ the following decomposed expression for the counter term:
\ba
I_{ct} \, \simeq \,  4\int d\theta \sqrt{\g} \left[-\frac{1}{2b^2}(e^{b\phi}-e^{-b\chi})\right] \, .
\label{I_ct2}
\ea

Now let us evaluate the on-shell action explicitly. 
It is straightforward to confirm that in the $b\to 0$ limit, the solution to Liouville fields with the Dirichlet boundary condition is written as 
\footnote{
This solution corresponds to (\ref{regular solution}) with $c^2 = 1-2b^2A$.
}
\ba
&& \phi \, = \, \frac{1}{b}\log\left(\frac{2}{1-r^2}\right) \, - \, Ab \, \frac{1+r^2}{1-r^2} \, , \\
&& \chi \, = \, -\frac{1}{b}\log\left(\frac{2}{1-r^2}\right) \, - \, Ab \, \frac{1+r^2}{1-r^2} \, ,  
\label{rone}
\ea
which is equivalent to the AdS$_2$ solution (\ref{adst}) i.e.
\ba
e^{2\rho} \, = \, \frac{4}{(1-r^2)^2} \, , \qquad \
\Phi \, = \, A \, \frac{1+r^2}{1-r^2} \, .
\ea
Then we can compute each on-shell action as 
	\begin{align}
        I^{\bulk}_{\rm L} \, &= \, - \, 8A\int d\theta \, \frac{r^2(1+r^2)}{(1-r^2)^2} \, , \\
        I^{\bdy}_{\rm L} \, &= \, -4A \int d\theta \, \frac{1+r^2}{1-r^2} \, , \\
        I_{\ct} \, &= \, 8A\int d\theta \, \frac{r(1+r^2)}{(1-r^2)^2} \, ,
	\end{align}
and by summing them we obtain 
\ba
I^{\bulk}_{\rm L}+I^{\bdy}_{\rm L}+I_{ct} \, = \, -4A\int d\theta \, \frac{1+r^2}{(1+r)^2} \, .
\ea
This indeed agrees with the JT on-shell action (\ref{jtac}).

%\begin{figure}
% \centering
%  \includegraphics[width=10cm]{AdSBCFT.pdf}
%  \caption{A sketch of gravity duals of CFT on a cylinder in the connected phase (left) and disconnected %phase (right).}
%\label{AdSBCFTfig}
%\end{figure}

%%%%%%%%%%%%%%%%%%%%%%%%%%%%%%%%%%%%%%%%%%%%%%%%%%%%%%%%%%%%%%%%%%%%%%
\subsection{FZZT brane interpretation}
\label{sec:FZZT brane}
%%%%%%%%%%%%%%%%%%%%%%%%%%%%%%%%%%%%%%%%%%%%%%%%%%%%%%%%%%%%%%%%%%%%%%

In the previous description  (\ref{rone}), we put the UV cut off $r=1-\delta$ by hand. 
Instead, it is also useful to regularize the solution itself with the boundary condition $\phi_0=\phi(r=1)$ and $\chi_0=\chi(r=1)$. 
This is expected to correspond to the semi-classical limit of FZZT-brane.\footnote{More precisely this type of boundary condition is the Legendre transform of the FZZT-brane boundary condition with a fixed boundary length.
Original FZZT-brane (at least classically) corresponds to a fixed energy boundary condition of JT gravity \cite{Goel:2020yxl}.
We will summarize boundary conditions of classical Liouville theory in appendix~\ref{app:BC's}.
}
We can find appropriate solutions by solving the Liouville equation as follows:
	\begin{align}
        e^{2b\phi} \, &= \, \frac{p_{\phi}^2 e^{2b\phi_0}}{(1-(1-p_{\phi})r^2)^2} \, , \no
        e^{-2b\chi} \, &= \, \frac{p_{\chi}^2 e^{-2b\chi_0}}{(1-(1-p_{\chi})r^2)^2} \, ,  
    \label{solpc}
	\end{align}
where $p_{\phi}$ and $p_{\chi}$ satisfy $p_{\phi}^2e^{2b\phi_0}=4(1-p_{\phi})$ and $p_{\chi}^2e^{-2b\chi_0}=4(1-p_{\chi})$.
Explicitly we find
	\begin{align}
        p_{\phi} \, &= \, -2e^{-2b\phi_0}+2e^{-b\phi_0}\s{1+e^{-2b\phi_0}} \, \simeq \, 2e^{-b\phi_0}-2e^{-2b\phi_0} \, > \, 0 \, ,\\
        p_{\chi} \, &= \, -2e^{2b\chi_0}+2e^{b\chi_0}\s{1+e^{2b\chi_0}} \, \simeq \, 2e^{b\chi_0}-2e^{2b\chi_0} \, > \, 0 \, .
	\end{align}

We are interested in the limit $b\to 0$ with 
\ba
&& \delta=e^{-\frac{b}{2}(\phi_0-\chi_0)}\ll 1,  \\
&&\eta\equiv b(\phi_0+\chi_0)=-\frac{2b^2A}{\delta}\ll  1,
\label{delta-eta}
\ea
while we require $\Phi_0=-\frac{1}{2b}(\phi_0+\chi_0)\gg 1$.
%
%In this limit, we can approximate the solution (\ref{solpc}) as follows
%\ba
%&& e^{2b\phi}=\frac{a^2(1+\eta)}{(1-(1-a\delta)r^2)^2},\\
%&& e^{-2b\chi}=\frac{\ti{a}^2 (1-\eta)}{(1-(1-\ti{a}\delta)r^2)^2},  \label{solpa}
%\ea
%where we introduced 
%\ba
%&& a=2(1-\delta)(1-\eta/2+\eta\delta/2),\no
%&& \ti{a}=2(1-\delta)(1+\eta/2-\eta\delta/2).
%\ea
%
Then the on-shell action is evaluated as 
\ba
&& I^{bulk}_L\simeq \int d\theta \left[\eta\left(\frac{2}{b^2\delta}-\frac{2}{b^2}+\frac{\delta}{b^2}\right)\right]\simeq A\int d\theta \left[-\frac{4}{\delta^2}+\frac{4}{\delta}-2\right],\\
&&I^{bdy}_L\simeq \frac{2}{b}\int d\theta (\phi_0+\chi_0)\simeq -\frac{4A}{\delta}\int d\theta.
\ea
The counter term is evaluated as 
\ba
 I_{ct}=4\int d\theta \left[-\frac{1}{2b}(\phi+\chi)e^{\frac{b}{2}(\phi-\chi)}\right] 
\simeq \frac{4}{\delta}\int d\theta \left(-\frac{1}{2b}(\phi_0+\chi_0)\right)
\simeq \frac{4A}{\delta^2}\int d\theta.
\ea
Therefore the total on-shell action becomes
\ba
I^{bulk}_L+I^{bdy}_L+I_{ct}=-2A\int d\theta=-4\pi A,
\label{I_on-shell}
\ea
which reproduces the JT gravity result (\ref{jtac}).

It is also useful to evaluate the contributions from $\phi$ and $\chi$ sector, separately.
First we note
\ba
&& \phi_0=-\frac{1}{b}\log\delta +\frac{\eta}{2b},\\
&& \chi_0=\frac{1}{b}\log\delta+\frac{\eta}{2b}.
\ea
Then, the evaluation of the on-shell action of each sector (detail evaluations are presented in appendix~\ref{app:on-shell action}) is given by
	\begin{align}
       I^{\scriptsize\mbox{on-shell}}_\phi \, = \, \int d\theta \left[-A+\frac{2}{b^2}(\log 2 -1) {  - \frac{\delta}{b^2}} \right] \, ,
    \label{I_phi}
	\end{align}
and
	\begin{align}
        I^{\scriptsize\mbox{on-shell}}_\chi \, = \, \int d\theta \left[-A-\frac{2}{b^2}(\log 2 -1) + \frac{\delta}{b^2} \right] \, .
    \label{I_chi}
	\end{align}
The total of two sectors agrees with the above result (\ref{I_on-shell}).

%%%%%%%%%%%%%%%%%%%%%%%%%%%%%%%%%%%%%%%%%%%%%%%%%%%%%%%%%%%%%%%%%%%%%%
\subsection{Minisuperspace wavefunctions}
\label{sec:minisuperspace}
%%%%%%%%%%%%%%%%%%%%%%%%%%%%%%%%%%%%%%%%%%%%%%%%%%%%%%%%%%%%%%%%%%%%%%

We can also derive the disk partition function from the minisuperspace wave functions
in the $b\to 0$ limit. We follow the minisuperspace analysis in \cite{Fateev:2000ik},
where the Liouville action 
looks like $I^{FZZ}_L=\frac{1}{4\pi}\int dx^2 [(\de_a\phi)^2+4\pi\mu^{FZZ} e^{2b\phi}]$
and it is $\frac{1}{4\pi}$ times our original action.
In our convention of $\mu(=4\pi \mu^{FZZ})$, we set $\mu b^2=1$ as before. This leads to 
\ba
\kappa \, = \, \s{\frac{\mu^{FZZ}}{\sin\pi b^2}} \, \simeq \, \frac{1}{2\pi b^2} \, .
\ea 
In \cite{Fateev:2000ik}, the bulk one-point function with a fixed boundary length was obtained as
	\begin{align}
        W^{(\phi)}_\a(l_\phi) \, &\equiv \, \big\la e^{2\a \phi} \big\lb_{l_\phi} \, = \, \frac{2}{b} \left( \pi \mu \gamma(b^2) \right)^{\frac{Q-2\alpha}{2b}} \, 
        \frac{\Gamma(2\alpha b - b^2)}{\Gamma(1+\frac{1}{b^2}-\frac{2\alpha}{b})} \, K_{\frac{Q-2\a}{b}}(\kappa l_\phi) \, , \label{W^phi} \\[4pt]
        W^{(\chi)}_\a(l_\chi) \, &\equiv \, \big\la e^{2\a \chi} \big\lb_{l_\chi} \, = \, \frac{2i}{b} \left( - \pi \mu \gamma(-b^2) \right)^{\frac{-q+2\alpha}{2b}} \, 
        \frac{\Gamma(2\alpha b + b^2)}{\Gamma(1-\frac{1}{b^2}+\frac{2\alpha}{b})} \, K_{\frac{-q+2\a}{b}}(-\kappa l_\chi) \, , \label{W^chi}
	\end{align}
where $\gamma(x)=\Gamma(x)/\Gamma(1-x)$ and we defined the one-point function of the time-like theory by simple analytical continuation  
	\begin{align}
        \phi \, \to \, -i \chi \, \quad b\, \to \, -i b \, , \quad \a \, \to \, i \a \, , \quad Q \, \to \, i q \, , \quad \m \, \to \, - \m \, ,
	\end{align}
from the result of the space-like theory.

In contrast to \cite{Saad:2019lba, Mertens:2020hbs}, where they used the ZZ brane boundary condition for the time-like Liouville theory,
here we use the FZZT brane boundary condition for the time-like theory.
This is because the rewriting of the action, we discussed in section~\ref{sec:jt limit}, the time-like field appeared on equal footing with the space-like field.
Furthermore, in the JT gravity metric (\ref{conformal gauge}), the physical boundary length is measured by $e^{2\r} = e^{b(\p - \c)}$.
Therefore, in order to compare with the JT gravity result, we should use this vertex operator $e^{b(\p - \c)}$.
Since the conformal dimension of the vertex operator $e^{b (\phi-\chi)}$ is one, 
in the minisuperspace approximation for $b\to 0$ limit, we expect that the disk amplitude with the Dirichlet boundary conditions 
$\phi=\phi_0$ and $\chi=\chi_0$ reads 
\ba
\big\la e^{b (\phi_0-\chi_0)} \big\lb_{\rm disk} \, \sim \, K_{\frac{1}{b^2}}(\k l_\phi)K_{\frac{1}{b^2}}(-\k l_\chi) \, ,
\ea
 where we set
	\begin{align}
        l_\phi \, &\equiv \, 2\pi e^{b\phi_0} \, \simeq \, 2\pi\cdot \delta^{-1}(1+\eta/2) \, ,\no
        l_\chi \, &\equiv \, 2\pi e^{-b\chi_0} \, \simeq \, 2\pi\cdot \delta^{-1}(1-\eta/2) \, .
	\end{align}
By using the approximation formula 
\ba
K_\nu (z) \, \simeq \, e^{\frac{\nu^2}{2z}-z} \, ,
\ea
for $\nu\gg 1$ and $z\gg \nu$,
we can estimate the disk amplitude as 
\ba
\big\la e^{b (\phi_0-\chi_0)} \big\lb_{\rm disk} \, \sim \, e^{\frac{\pi}{b^2 l_\phi}-\frac{\pi}{b^2 l_\chi}} \, \simeq \, e^A \, .
\ea
By multiplying the $4\pi$ factor to adjust to our original convention this leads to 
\ba
\big\la e^{b (\phi_0-\chi_0)} \big\lb_{\rm disk} \, \sim \, e^{4\pi A} \, .
\ea
This indeed agrees with the JT gravity calculation (\ref{jtac}).

We can also consider each contribution separately.
In order to agree with (\ref{I_phi}) and (\ref{I_chi}), we need to include precise coefficients as in (\ref{W^phi}) and (\ref{W^chi}).
Since 
\ba
&&K_{\frac{1}{b^2}}(\kappa l_\phi) \, \sim \, e^{\frac{\pi}{b^2 l_\phi}} \, \approx \, e^{\frac{\delta}{2b^2}\left( 1-\frac{\eta}{2} \right)} \, , \\[4pt]
&&K_{\frac{1}{b^2}}(-\kappa l_\chi) \, \sim \, e^{-\frac{\pi}{b^2 l_\chi}} \, \approx \, e^{- \frac{\delta}{2b^2}\left( 1+\frac{\eta}{2} \right)} \, ,
\ea
we find the contribution from the spacelike and timelike sector as 
	\begin{align}
W^{(\phi)}_\frac{b}{2}(l_\phi) \, & \simeq \, \exp\left[ - \frac{1}{2}\left( \frac{\delta \eta}{2b^2} - \frac{\delta}{b^2} + \frac{2}{b^2}(\log2-1) \right) \right] \\[4pt]
W^{(\chi)}_{-\frac{b}{2}}(l_\phi) \, & \simeq \, \exp\left[ - \frac{1}{2}\left( \frac{\delta \eta}{2b^2} + \frac{\delta}{b^2} - \frac{2}{b^2}(\log2-1) \right) \right]
	\end{align}
respectively.
These agree with (\ref{I_phi}) and (\ref{I_chi}) once we multiply the $4\pi$ factor as above.

%%%%%%%%%%%%%%%%%%%%%%%%%%%%%%%%%%%%%%%%%%%%%%%%%%%%%%%%%%%%%%%%%%%%%%
\subsection{Annulus amplitude}
\label{sec:annulus amplitude}
%%%%%%%%%%%%%%%%%%%%%%%%%%%%%%%%%%%%%%%%%%%%%%%%%%%%%%%%%%%%%%%%%%%%%%
In this section, we would like to study annulus amplitude by semiclassical approximation.
To this end, we again take the route we used for JT gravity in section~\ref{JTwormhole}.
Namely, we study a bulk one-point function $\big\la e^{2\pi \a (\p-\c)} \big\lb$ inserted at the center of disk.
The corresponding semiclassical solutions are \cite{Seiberg:1990eb}
	\begin{align}
		e^{2b\phi} \, &= \, \frac{p_{\phi}^2 e^{2b\phi_0} \, r^{2\n-2}}{(1-(1-p_{\phi})r^{2\n})^2} \, , \no
		e^{-2b\chi} \, &= \, \frac{p_{\chi}^2 e^{-2b\chi_0} \, r^{2\n-2}}{(1-(1-p_{\chi})r^{2\n})^2} \, ,  
	\end{align}
where $\n$ is defined in (\ref{nu}) and now 
	\begin{align}
		p_{\phi}^2e^{2b\phi_0} \, &= \, 4 \n^2 (1-p_{\phi}) \, , \no
		p_{\chi}^2e^{-2b\chi_0} \, &= \, 4 \n^2 (1-p_{\chi}) \, .
	\end{align}
We use the same parametrization of $\d$ and $\h$ as before (\ref{delta-eta}).	
We present detail evaluation of the on-shell action in appendix~\ref{app:on-shell action} and simply quote the final result here:
	\begin{align}
		I^{bulk}_L+I^{bdy}_L+I_{ct} \, = \, -2\n^2 A\int d\theta \, = \, -4\pi \n^2 A \, .
	\end{align}
This agrees with the JT gravity result (\ref{JT-pdisk}).
Then, the tree-level annulus amplitude is now given by
	\begin{align}
		\int_0^{\inf} d\tilde{\a} \, \tilde{\a} \, e^{- 2\tilde{\a}^2 \frac{\bar{\Phi}}{\beta_1}} e^{- 2\tilde{\a}^2 \frac{\bar{\Phi}}{\beta_2}} \, = \, \frac{\b_1 \b_2}{4\bar{\Phi}(\b_1 +\b_2)} \, .
	\end{align}

We can also recover these results from the minisuperspace wavefunctions:
	\begin{align}
		W_P(l) \, = \, \big( \pi \mu \g(b^2) \big)^{-\frac{iP}{b}} \, \frac{\Gamma(1+2ibP)}{\Gamma(-\frac{2iP}{b})} \, K_{\frac{2i P}{b}} (\k l) \, .
	\end{align}
For spacelike theory, we substitute $2iP=\n (b - b^{-1})$ with $\k l= \k l_\phi$, while for timelike theory we use $2iP=\n (b + b^{-1})$ with $\k l= -\k l_\chi$.
This leads to the contribution from the spacelike and timelike sector
	\begin{align}
		&\exp\left[ - \frac{1}{2b^2}\left( 2\n \log 2\n - 2\n + \n^2 \d - \frac{\n^2 \d \h}{2} \right) \right] \, , \\[4pt]
		&\exp\left[ - \frac{1}{2b^2}\left( -2\n \log 2\n + 2\n - \n^2 \d - \frac{\n^2 \d \h}{2} \right) \right] \, ,
	\end{align}
respectively.

%%%%%%%%%%%%%%%%%%%%%%%%%%%%%%%%%%%%%%%%%%%%%%%%%%%%%%%%%%%%%%%%%%%%%%
\subsection{Boundary Schwarzian theory}
\label{sec:Schwarzian theory}
%%%%%%%%%%%%%%%%%%%%%%%%%%%%%%%%%%%%%%%%%%%%%%%%%%%%%%%%%%%%%%%%%%%%%%
In this subsection, we study the boundary theory of our double Liouville theory defined in section~\ref{sec:liouville action}.
For this discussion, it's more convenient to use complex coordinates: $z=re^{i \th}$ and $\bar{z}=r e^{-i \th}$.
Therefore, now the metric is given by $ds^2 =dz d\bar{z}$ and the bulk action becomes
	\begin{align}
        I^{bulk}_L \, = \, \int dz d\bar{z} \Big[ 4\pa \phi \bar{\pa} \phi \, - \, 4\pa \chi \bar{\pa} \chi \, + \, \mu (e^{2b\phi} - e^{-2b\chi}) \Big] \, .
	\end{align}
This action is invariant under the conformal transformation
	\begin{align}
        z \, \to f(z) \, , \qquad \bar{z} \, \to \, \bar{f}(\bar{z}) \, , 
	\label{diffeo}
	\end{align}
together with
	\begin{align}
        \phi(z, \bar{z}) \, &\to \, \phi(f(z), \bar{f}(\bar{z})) \, - \, \frac{1}{2b} \, \log(\pa f \bar{\pa} \bar{f}) \, , \\[2pt]
        \chi(z, \bar{z}) \, &\to \, \chi(f(z), \bar{f}(\bar{z})) \, + \, \frac{1}{2b} \, \log(\pa f \bar{\pa} \bar{f}) \, .
	\end{align}
We note that the bulk action does not have independent transformation for each Liouville theory, because this is a coordinate transformation (\ref{diffeo}).

At the boundary, the holomorphic and anti-holomorphic functions are related by the condition $f(\bar{z}) = \bar{f}(z)$.
This means that at the boundary, we have the diffeomorphism of the angular coordinate as $\th \to f(\th)$.
This symmetry is explicitly broken by the boundary term (\ref{LVbd}).
Taking $b\to0$ limit and combining with the counterterm (\ref{I_ct1}), this boundary action is
	\begin{align}
        I_L^{\bdy+\ct} \, = \, \frac{2}{b} \int d\th \, (\phi+\chi) ( K^{(0)} - 1 ) \, .
	\label{I_bdy+ct}
	\end{align}
In order to evaluate this action, we can again go back to the polar coordinates and the boundary condition gives $r(\th) = 1 - \e f'(\th) $ as in (\ref{r(u)}).
This leads to 
	\begin{align}
        &\phi+\chi \, = \, - \frac{2Ab}{\e} \, + \, \mathcal{O}(\e^0) \, , \\ 
        &K^{(0)} \, = \, 1 \, + \, \left( \Sch\big( \tan(\tfrac{f}{2}) \, , \th \big) \, + \, \frac{1}{2} \right) \e \, + \, \mathcal{O}(\e^2) \, .
    \label{K^0-expansion}
	\end{align}
Therefore, at the order $\e^0$, we find the Schwarzain action up to a constant as
	\begin{align}
        I_\Sch \, = \, -4A \int_0^{2\pi} d\th \, \Sch\big( \tan(\tfrac{f}{2}) \, , \th \big) \, .
	\label{I_Sch-Lv}
	\end{align}
This agrees with (\ref{I_Sch}).
Since all results of JT gravity with a fixed background topology is obtained by the Schwarzain theory,
this confirms that our double Liouville theory indeed gives JT gravity results in the semiclassical limit ($b \to 0$).

\section{$c=1$ Matrix Model Interpretation}
\label{sec:matrix model}
%%%%%%%%%%%%%%%%%%%%%%%%%%%%%%%%%%%%%%%%%%%%%%%%%%%%%%%%%%%%%%%%%%%%%%
The CFT defined by a space-like Liouville CFT coupled to a 
free scalar $c=1$ CFT, such that total central charge is $26$, defines a world-sheet theory of two dimensional string theory, so called $c=1$ string theory. It is well-known that this string theory is dual to the $c=1$ matrix model \cite{ Klebanov:1991qa,Ginsparg:1993is,Polchinski:1994mb}.

A $c<1$ extension of two dimensional string theory can be obtained by considering a space-like Liouville CFT coupled to a time-like Liouville theory as in  (\ref{LVb}).  This string theory can be regarded as a non-critical string theory with  $c\leq 1$ non-rational matter. Its matrix model description was given in \cite{Takayanagi:2004yr} by deforming the $c=1$ matrix model state.
As a special case, this includes the familiar $c=1$ string theory. 
To relate this to the JT gravity, we are interested in the limit $b\to 0$, which means that the matter central charge is $c=-\infty$. 
\footnote{The $c=1$ matrix model in this context has also recently been discussed in \cite{Betzios:2020nry}.}
In addition to the ones discussed in \cite{Saad:2019lba, Mertens:2020hbs, Johnson:2019eik}, 
this matrix model gives yet another non-perturbative definition of JT gravity.
It would be interesting to study the difference of these non-perturbative completion of JT gravity more in detail. 
Here, we discuss our initial attempts towards this direction.

%%%%%%%%%%%%%%%%%%%%%%%%%%%%%%%%%%%%%%%%%%%%%%%%%%%%%%%%%%%%%%%%%%%%%%
\subsection{$c=1$ Matrix model and deformation}
\label{sec:matrix modelf}
%%%%%%%%%%%%%%%%%%%%%%%%%%%%%%%%%%%%%%%%%%%%%%%%%%%%%%%%%%%%%%%%%%%%%%

The $c=1$ matrix \cite{Gross:1990ay, Brezin:1989ss, Ginsparg:1990as} is defined by the following matrix quantum mechanics:
\ba
S_M=\int dt \, \Tr\left[(D_{t}\Phi)^2+\Phi^2\right],
\ea
where $\Phi(t)$ is a $N\times N$ hermitian matrix and $D_t \bullet=\de_t \bullet -i[A_t,\bullet]$ is the covariant 
derivative, so that the action is invariant under $U(N)$ symmetry. By using this $U(N)$ gauge symmetry,
we can diagonalize the matrix $\Phi$ into $N$ eigenvalues. They are identical to $N$ free fermions, each of which is described by the Hamiltonian with the inverse harmonic potential $H=p^2-x^2$.

In the $N\to \infty$ limit, the fermions fills the energy level up to the Fermi surface 
\ba
H=p^2-x^2=-\mu,
\ea
where we only fill the left-hand side of the potential $x<0$.
This defines the vacuum state of $c=1$ matrix model. 

The $c\leq 1$ string theory is defined by the world-sheet theory 
\ba
I^{c<1}=\int dx^2 \left[(\de_a\phi)^2-(\de_a\chi)^2+\mu e^{2b\phi}+\nu e^{-2b\chi} 
\right]. \label{LVnon}
\ea
This can be boosted 
into a deformation of $c=1$ string via the map:
	\begin{align}
        \ti{\phi} \, &= \, \frac{Q}{2}\phi+\frac{q}{2}\chi,\no
        \ti{\chi} \, &= \, \frac{q}{2}\phi+\frac{Q}{2}\chi,
	\end{align}
which leads to 
\ba
I^{c<1}=\int dx^2 \left[(\de_a\ti{\phi})^2-(\de_a\ti{\chi})^2+\mu e^{(b^2+1)\ti{\phi}+(b^2-1)\chi}
+\nu e^{(1-b^2)\ti{\phi}-(1+b^2)\ti{\chi}} 
\right]. \label{LVnona}
\ea

\begin{figure}[t!]
	\begin{center}
        \includegraphics[width=6cm]{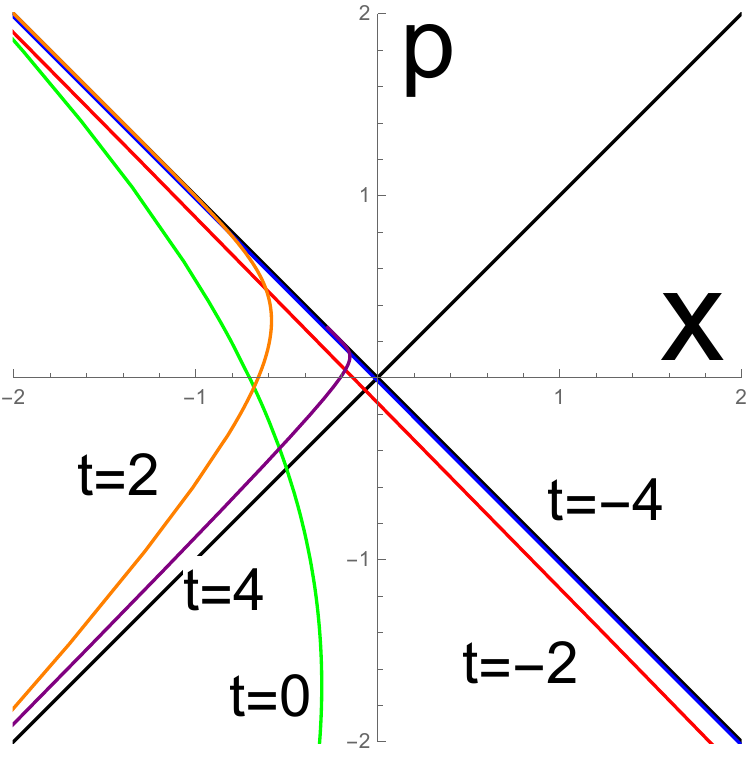}
	\end{center}
    \caption{Profiles of Fermi surfaces for $t=-4$ (blue), $t=-2$ (red) $t=0$ (green), $t=2$ (orange) and $t=4$ (purple).
    The Fermi seas extend in the left-hand side of the Fermi surfaces.}
\label{FSfig}
\end{figure}

This background (\ref{LVnona}) corresponds to a series of excited states in the $c=1$ matrix model 
given by the following Fermi surface \cite{Takayanagi:2004yr} : 
\ba
(p-x)(-p-x)^{b^2}=\mu e^{-(1-b^2)t}+\nu (-p-x)^{2b^2} e^{-(1+b^2)t}.
\label{fermi surface}
\ea
In our case of the $b\to 0$ limit with $\mu =-\nu=1/b^2$, the Fermi surface takes the following characteristic form:
\ba
(p-x)e^{t}=-2\log\left[(-p-x)e^{-t}\right].  \label{FSN}
\ea
This is plotted in Fig.\ref{FSfig}. At $t=-\infty$ we start with the Fermi surface $p+x<0$ and eventually it approaches to the wedge $p+x<0$ and $p-x>0$ in the late limit $t\to \infty$.

%%%%%%%%%%%%%%%%%%%%%%%%%%%%%%%%%%%%%%%%%%%%%%%%%%%%%%%%%%%%%%%%%%%%%%
\subsection{Collective field description}
\label{sec:matrix modelc}
%%%%%%%%%%%%%%%%%%%%%%%%%%%%%%%%%%%%%%%%%%%%%%%%%%%%%%%%%%%%%%%%%%%%%%

Next we study the effective theory which captures excitations  on our Fermi surface (\ref{FSN})
by employing the collective field theory \cite{Das:1990kaa,Jevicki:1993qn} (see also \cite{Polchinski:1991uq,Alexandrov:2003uh}).
We introduce the collective field by density of eigenvalues \cite{Jevicki:1979mb} as
	\begin{align}
        \vp(t, x) \, \equiv \, \sum_{i=1}^N \d(x - \l_i(t)) \, ,
	\label{collective field}
	\end{align}
where $\l_i(t)$ is an eigenvalue of the matrix $\Phi(t)$.
The collective field after Fourier transforming from $x$ to $k$ corresponds to the gauge invariant (Wilson) loop operators
	\begin{align}
        \vp_k(t) \, = \, \int dx \, e^{ikx} \vp(t, x) \, = \, \sum_{i=1}^N e^{ik \l_i} \, = \, \Tr\Big( e^{ik \Phi(t)}\Big) \, .
	\end{align}
In terms of this collective field, the matrix quantum mechanics is described by 
	\begin{align}
        S_{\rm coll} \, = \, \int dt dx \, \left( \frac{1}{\vp} \left( \int dx \, \pa_t \vp \right)^2 - \frac{\pi^2}{3} \vp^3 + (x^2 -\m) \vp \right) \, .
	\end{align}

When the perturbation of the Fermi surface is infinitesimally small,
this is described by small fluctuations of the collective field around the background solution:
	\begin{align}
        \vp(t,x) \, = \, \vp_0(t,x) \, + \, \frac{1}{\sqrt{\pi}} \, \pa_x \h(t,x) \, , 
	\end{align}
where
	\begin{align}
        \vp_0(t,x) \, = \, \frac{1}{2\pi} \, (p_+ - p_-) \, .
	\end{align}
Then, the quadratic action of the fluctuation $\eta$ becomes that of a real massless scalar with the kinetic term \cite{Das:1990kaa,Jevicki:1993qn}:
\ba
I_{sc}=-\frac{1}{2}\int dt dx \s{g}g^{\mu\nu}\de_\mu \eta\de_\nu\eta,
\ea
where the effective metric $g_{\mu\nu}$ is defined by 
\ba
ds^2&=&g_{\mu\nu}dx^\mu dx^\nu=\frac{2p_+p_-}{p_+-p_-}dt^2-\frac{2(p_++p_-)}{p_+-p_-}dtdx
+\frac{2}{p_+-p_-}dx^2.
\ea

Before we study the collective field dynamics of our background, it is helpful to quickly 
review the standard result for the usual $c=1$ vacuum, which is described by the Fermi surface $p^2-x^2=-1$.
In this case, we can express the Fermi surface by introducing a parameter $\sigma$ or $\ti{\sigma}$
which parameterizes the branch $p=p_+(x)$ or $p=p_-(x)$, respectively, as follows: 
\ba
&& (-p_+-x)e^{-t}=e^{\sigma},\ \ \ (p_+-x)e^t =e^{-\sigma},\no
&& (-p_--x)e^{-t}=e^{\ti{\sigma}},\ \ \ (p_--x)e^t =e^{-\ti{\sigma}},
\label{sigam-sigamtilde for c=1}
\ea
where $\sigma=-\ti{\sigma}-2t$ runs in the range 
\ba
&&-\infty< \sigma \leq \sigma_0(t),  \no
&&  \sigma_0(t)\leq \ti{\sigma}<\infty,  \label{rangef}
\ea
with $\sigma_0(t)=-t$.

\begin{figure}
 \centering
  \includegraphics[width=7cm]{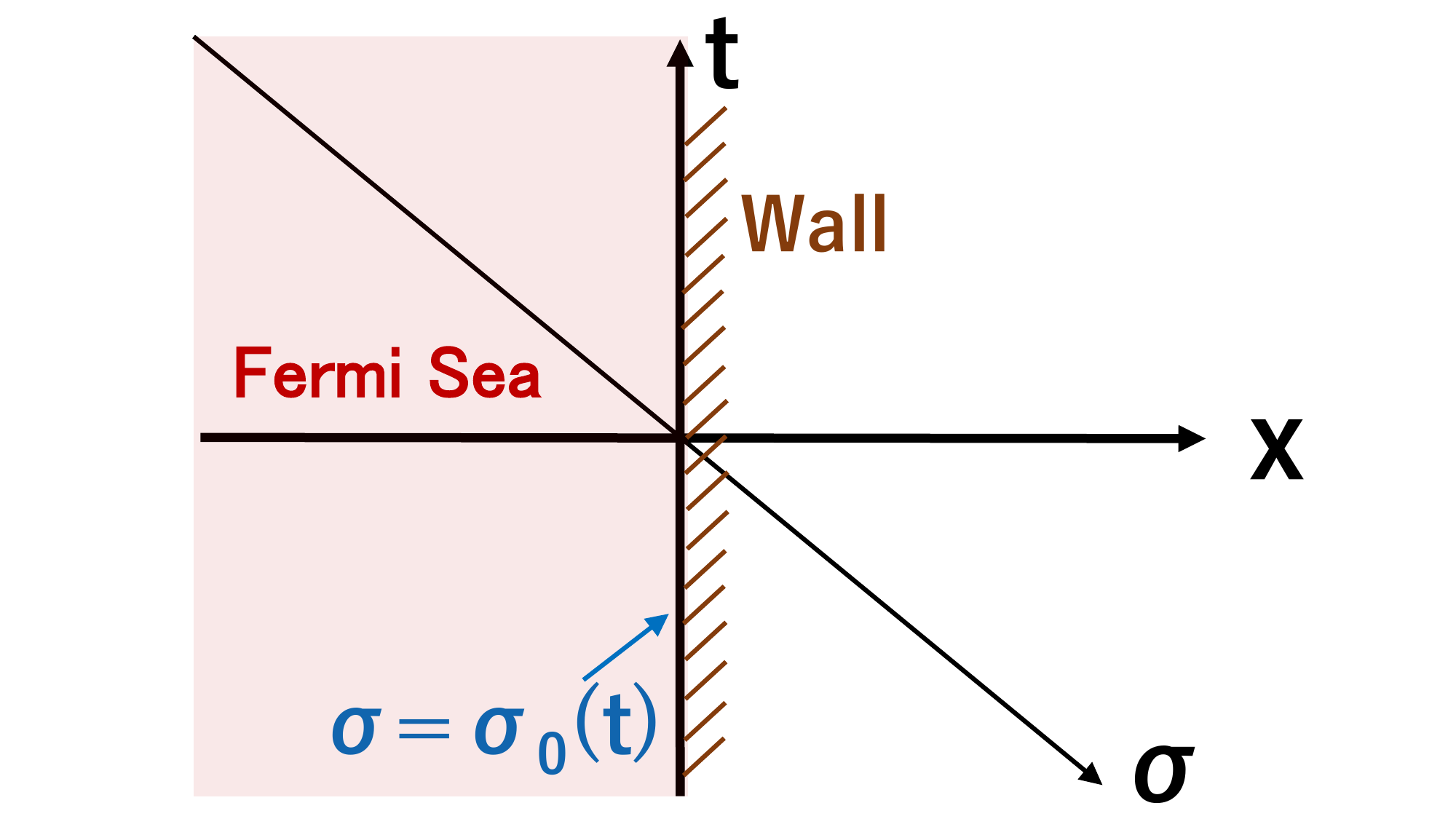}
  \includegraphics[width=7cm]{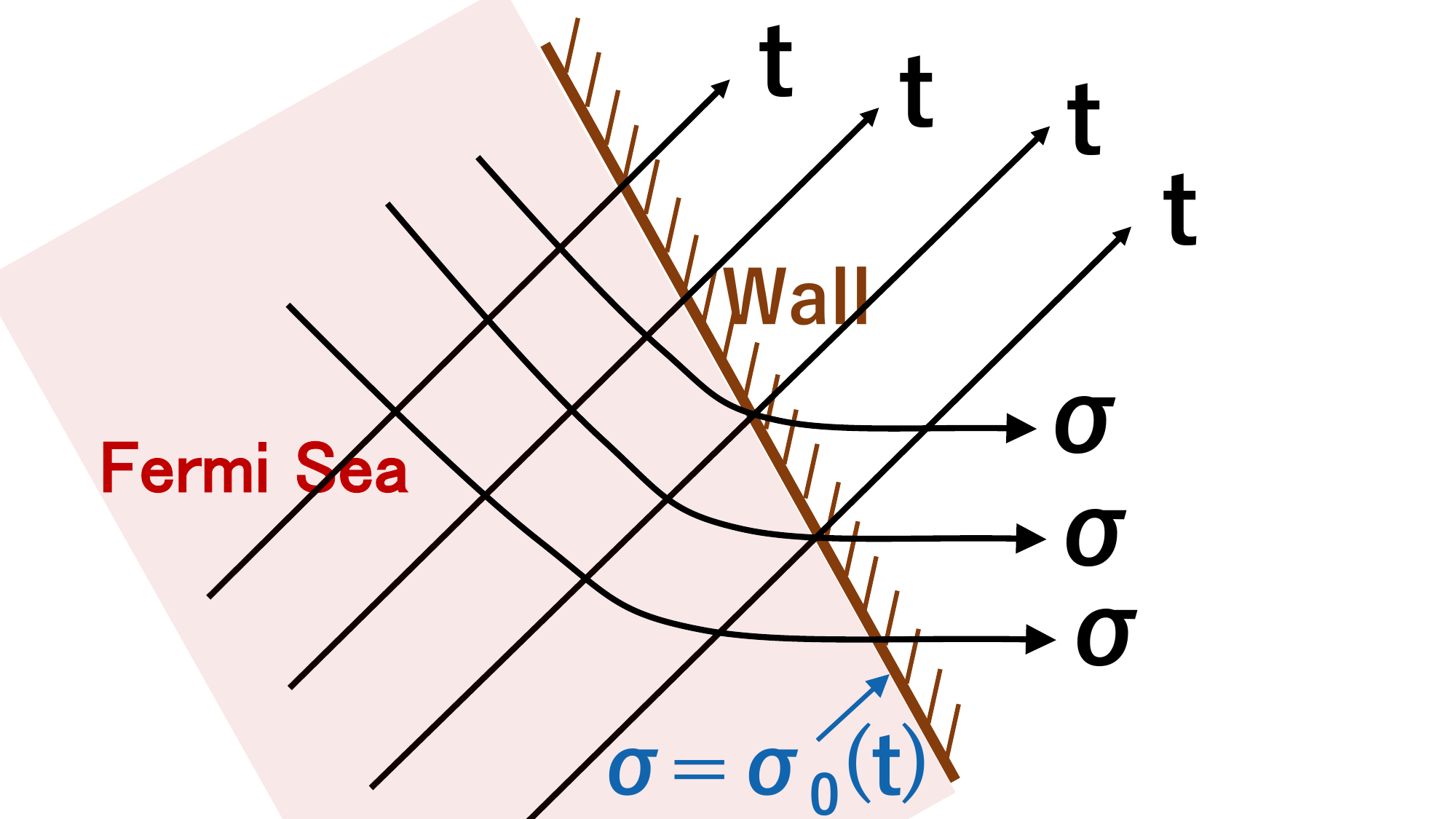}
 \caption{Structures of Spacetimes of the collective field theory.
 The left and right picture describe the standard $c=1$ vacuum and our background, described by (\ref{CFta}), respectively.} \label{STfigqq}
\end{figure}

Thus the metric (\ref{metgen}) reads
\ba
ds^2=\frac{1}{2}(e^{-t-\sigma}-e^{t+\sigma})\left[-dt^2+d(t+\sigma)^2\right].
\ea
If we introduce a new coordinate
\ba
q=\sigma+t,
\ea
then the effective action takes the standard form
\ba
I_{sc}=-\frac{1}{2}\int dt \int^0_{-\infty} dq \left[(\de_t\eta)^2-(\de_q\eta)^2\right].
\ea
This shows the well-known fact that the target spacetime geometry of $c=1$ matrix model is a half space $q\leq 0$ with a Liouville wall at $q=0$, as sketched in the left of Fig.\ref{STfigqq}.

\begin{figure}[t!]
	\begin{center}
       \includegraphics[width=10cm]{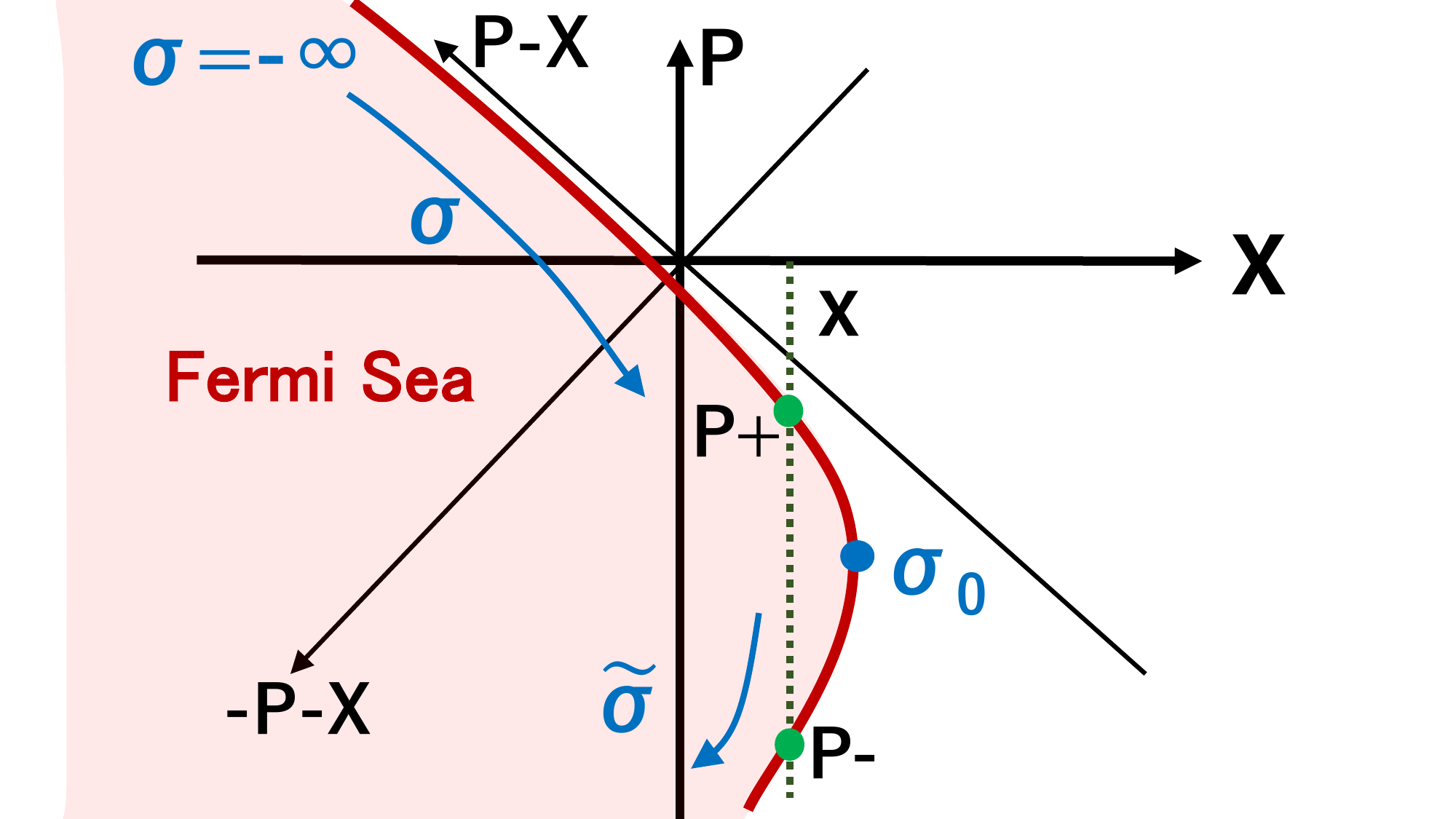}
	\end{center}
    \caption{Structure of Fermi surfaces paramerized by $\sigma$ and $\ti{\sigma}$. We also showed momentum functions $p_\pm(x)$.} 
\label{FSPfig}
\end{figure}

Now let us consider the collective field theory for our background defined by the time-dependent 
fermi surface (\ref{FSN}). The fixed $x$ line intersects with the Fermi surface at two points,
written by $p=p_+(x)$ and $p=p_-(x)$ as depicted in Fig.\ref{FSPfig}.
We introduce a new parameter $\sigma$ and $\ti{\sigma}$ instead of $x$, which parameterize the branches $p=p_+(x)$ and $p=p_-(x)$, respectively as:
\ba
&& (-p_+-x)e^{-t}=e^{\sigma},\ \ \ (p_+-x)e^t =-2\sigma,\no
&& (-p_--x)e^{-t}=e^{\ti{\sigma}},\ \ \ (p_--x)e^t =-2\ti{\sigma}.
\ea
The parameters $\sigma$ and $\ti{\sigma}$ take the values in the range (\ref{rangef}), where
we can explicitly find $\sigma_0(t)$, where the two branches get degenerate, as 
\ba
\sigma_0(t)=-2t+\log 2.
\ea

We can solve $x$ and $p_{\pm}$ as
\ba
&& x=-\frac{1}{2}e^{t+\sigma}+\sigma e^{-t}=-\frac{1}{2}e^{t+\ti{\sigma}}+\ti{\sigma}e^{-t},\no
&& p_+=-\frac{1}{2}e^{t+\sigma}-\sigma e^{-t},\no
&& p_-=-\frac{1}{2}e^{t+\ti{\sigma}}-\ti{\sigma}e^{-t}.
\ea
It is also useful to note the relations
\ba
p_+-p_-=2e^{-t}(\ti{\sigma}-\sigma),\ \ \ \ p_++p_-=-e^{t+\sigma}-2\ti{\sigma} e^{-t}.
\ea

In our back ground, the metric is explicitly found as
\ba
ds^2 =2\left(-\frac{1}{2}e^{t+\sigma}+e^{-t}\right)dt d\sigma+\frac{e^t}{\ti{\sigma}-\sigma}\left(-\frac{1}{2}e^{t+\sigma}+e^{-t}\right)^2
d\sigma^2.  \label{metgen}
\ea
However note that the metric is meaningful only up to Weyl rescaling due to the conformal invariance of the massless scalar.  This leads to the effective action
\ba
I_{sc}=\frac{1}{2}\int dt \int^{\sigma_0}_{-\infty}d\sigma\left[
\frac{1-\frac{1}{2}e^{2t+\sigma}}{\ti{\sigma}-\sigma}(\de_t\eta)^2-2(\de_t\eta)(\de_\sigma\eta)
\right].\label{CFta}
\ea

If we take the asymptotic limit given by $\sigma\to -\infty$ and $\ti{\sigma}\to \infty$,
this action is simply approximated by 
\ba
I_{sc}=\frac{1}{2}\int dt \int^{\sigma_0(t)}_{-\infty}d\sigma\left[-2(\de_t\eta)(\de_\sigma\eta)
\right],
\ea
which suggests that $t$ and $\sigma$ plays a role of null coordinate. 
The metric is approximated by 
\ba
ds^2\simeq 2e^{-t}dtd\sigma.
\ea

On the other had, in the opposite limit $\sigma\simeq \sigma_0$, the action is estimated as
\ba
I_{sc}=\frac{1}{2}\int dt \int^{\sigma_0(t)}_{-\infty}d\sigma\left[
\frac{1}{2}(\de_t\eta)^2-2(\de_t\eta)(\de_\sigma\eta)
\right],
\ea
which shows that $t$ is null and $\sigma$ is space-like. 
The metrix is approximated by 
\ba
ds^2\simeq (\sigma_0-\sigma)e^{-t}\left(2dtd\sigma +\frac{1}{2}d\sigma^2\right).
\ea

From the above analysis shows that this target spacetime geometry is given by a time-like Liouville wall, as sketched in the right of Fig.\ref{STfigqq}. Therefore, we expect that this is a regular time-dependent background in $c=1$ matrix.

%%%%%%%%%%%%%%%%%%%%%%%%%%%%%%%%%%%%%%%%%%%%%%%%%%%%%%%%%%%%%%%%%%%%%%
\section{Two-dimensional de Sitter Gravity}
\label{sec:ds2}
%%%%%%%%%%%%%%%%%%%%%%%%%%%%%%%%%%%%%%%%%%%%%%%%%%%%%%%%%%%%%%%%%%%%%%
In this section, let us briefly consider the Liouville CFT and the matrix quantum mechanics which correspond to the two-dimensional de Sitter JT gravity
\cite{Maldacena:2019cbz, Cotler:2019nbi}.
From the discussion of section~\ref{sec:jt limit}, it looks like we just need to change the sign of the bulk cosmological constant (i.e. $\m \to -\m$)
in order to get dS$_2$ JT gravity in the semicalassical limit. 
Since the bulk cosmological constant appears inside of the logarithm as in (\ref{regular solution}),
the change of the sign induces a pure imaginary factor $i$ inside of the log in (\ref{rone}), so that we now have
	\begin{align}
        \phi \, &= \, \frac{1}{2b}\log\left(\frac{-4}{(1-r^2)^2}\right) \, - \, Ab \, \frac{1+r^2}{1-r^2} \, , \\
        \chi \, &= \, -\frac{1}{2b}\log\left(\frac{-4}{(1-r^2)^2}\right) \, - \, Ab \, \frac{1+r^2}{1-r^2} \, .
	\end{align}
Then for the metric, we find the ``negative AdS$_2$'' as discussed in \cite{Maldacena:2019cbz}
	\begin{align}
        e^{2\rho} \, = \, - \, \frac{4}{(1-r^2)^2} \, , \qquad \
        \Phi \, = \, A \, \frac{1+r^2}{1-r^2} \, .
	\end{align}
This overall minus sign gives a $-i$ factor for the tree-level on-shell action as
	\begin{align}
        I^{bulk}_L+I^{bdy}_L+I_{ct} \, = \, 4\pi i A \, .
	\end{align}
For the Schwarzian action discussed in section~\ref{sec:Schwarzian theory}, again the overall minus sign of the metric gives an overall $-i$ factor.
Therefore, the wavefunction of the two-dimensional de Sitter universe $\Psi_+$ is given from the AdS partition function by \cite{Maldacena:2019cbz}
	\begin{align}
        \Psi_+ \, = \, Z_{{\rm AdS}_2}\big(A_{\rm AdS} \to - i A_{\rm dS}\big).
	\end{align}
In the random matrix viewpoint, the original disk partition function $Z_{{\rm AdS}_2}$ in AdS$_2$ is given by $\mbox{Tr}[e^{-L\hat{H}}]$, where $\hat{H}$ is the Hamiltonian and $L$ is the boundary length. This is transformed into the wave function $\Psi_+$  in dS$_2$ by replacing $L$ with $-iL$ as  $\mbox{Tr}[e^{iL\hat{H}}]$ as argued in \cite{Maldacena:2019cbz}. This implies that the matrix quantum mechanics, we discussed in section~\ref{sec:matrix modelf}, does not change,
except that we replace the boundary length parameter $L$ with $-iL$.
Indeed, we can confirm that the matrix model description is equivalent both for
AdS$_2$ and dS$_2$. This is because in terms of the Fermi surface of 
the matrix model (\ref{fermi surface}),
the change of the sign of the bulk cosmological constant corresponds to the exchange of $x$ and $p$. Therefore, the Fermi surface (see Fig.~\ref{FSfig}) now moves into $p-x<0$ quadrant as $t$ grows. However, this does not mean any physical change and indeed the both are related by a canonical transformation.

%%%%%%%%%%%%%%%%%%%%%%%%%%%%%%%%%%%%%%%%%%%%%%%%%%%%%%%%%%%%%%%%%%%%%%
\section{Conclusions}
\label{sec:conclusions}
%%%%%%%%%%%%%%%%%%%%%%%%%%%%%%%%%%%%%%%%%%%%%%%%%%%%%%%%%%%%%%%%%%%%%%

In this paper, we argued an equivalence between the JT gravity on an anti de-Sitter space
and the two dimensional string theory defined by a time-like Liouville CFT coupled to the space-like Liouville one. We confirmed that their actions, disk partition functions and annulus amplitudes perfectly agree with each other, where the presence of boundary terms plays an important role. We also 
reproduced the boundary Schwarzian theory from the Liouville 
theory description. Our two dimensional string theory looks different from the minimal string theory in that the latter assumes the truncation to a rational CFT, while the former does not. Nevertheless, as we explicitly worked out in this paper, our two dimensional string theory which is equivalent to the JT gravity, has a non-perturbative description in terms of a time-dependent background in the $c=1$ matrix quantum mechanics. It will be an important future problem to prove the proposed correspondence between the JT gravity and the matrix quantum mechanics by explicitly evaluating the disk amplitudes of the latter. For this we need to precisely identify the form of loop operator in the matrix quantum mechanics, which fits with our $c<1$ two dimensional string theory. It will be also intriguing to explore what we can learn about holography in de-Sitter space

%%%%%%%%%%%%%%%%%%%%%%%%%%%%%%%%%%%%%%%%%%%%%%%%%%%%%%%%%%%%%%%%%%%%%%
\section*{Acknowledgements}
%%%%%%%%%%%%%%%%%%%%%%%%%%%%%%%%%%%%%%%%%%%%%%%%%%%%%%%%%%%%%%%%%%%%%%

We are grateful to Douglas Stanford for comments on the draft and Tomonori Ugajin for useful discussions. We would like to thank Gaston Gribet for a useful correnpondence.
This work is supported by the Simons Foundation through the ``It from Qubit'' collaboration.
TT is also supported by Inamori Research Institute for Science and World Premier International Research Center Initiative (WPI Initiative)
from the Japan Ministry of Education, Culture, Sports, Science and Technology (MEXT),
by JSPS Grant-in-Aid for Scientific Research (A) No.~21H04469 and
by JSPS Grant-in-Aid for Challenging Research (Exploratory) 18K18766.

\appendix
%%%%%%%%%%%%%%%%%%%%%%%%%%%%%%%%%%%%%%%%%%%%%%%%%%%%%%%%%%%%%%%%%%%%%%
\section{Boundary $\r K^{(0)}$ contribution}
\label{app:rho K0}
%%%%%%%%%%%%%%%%%%%%%%%%%%%%%%%%%%%%%%%%%%%%%%%%%%%%%%%%%%%%%%%%%%%%%%
As we commented in the footnote~\ref{foot:rho K0}, if we use (\ref{rho&Phi}) directly in (\ref{LVbd}), what we obtain is
	\begin{align}
	    I^{\bdy}_{\rm L} \, = \, 4\int d\theta \, (\rho - \Phi) K^{(0)} \, .
	\end{align}
Therefore, we find an additional term containing $\rho K^{(0)}$ in the boundary action.
In this appendix we make some comments on the effects of this additional term.

First let us consider the contribution of this term to the on-shell action
	\begin{align}
	    I^{\bdy}_{\rm L} \, \supset \, 4\int d\theta \, \rho K^{(0)} \, \simeq \, 2b \int d\th \, (\phi_0 - \chi_0) \, \simeq \, -8\pi \log \d \, .
	\end{align}
This is diverging, so it does not give any finite contribution to the on-shell action,
but we need an additional counterterm to eliminate this diverging contribution.
The additional counterterm can be taken as the following expression 
	\begin{align}
	    I^{\ct}_{\rm extra} \, = \, - 4\int d\theta \, \rho \, \simeq \, 8\pi \log \d \, ,
	\end{align}
which indeed eliminates the diverging contribution coming from the additional boundary term.

Next, we consider the derivation of the Schwarzian action discussed in section \ref{sec:Schwarzian theory}.
If we include the additional boundary and counter terms, instead of (\ref{I_bdy+ct}), we now obtain
	\begin{align}
        I_{\rm L}^{\bdy+\ct} \, = \, 4\int d\theta \, (\rho - \Phi) ( K^{(0)} - 1 ) \, .
	\end{align}
Using the expansion $r(\th) = 1- \ep f'(\th)$, we find
	\begin{align}
        \rho \, = \, \frac{b}{2} (\phi - \chi) \, = \, - \log \ep \, + \, \mathcal{O}(\ep) \, .
	\end{align}
Combining with (\ref{K^0-expansion}), we can see that the additional boundary term does not lead to any finite contribution in the $\ep \to 0$ limit.
Therefore, even if we include the additional boundary term, the effective action is still given by the Schwarzian action (\ref{I_Sch-Lv}).

It might be interesting to further clarify this discrepancy between the double Liouville theory and JT gravity,
but we leave this study to future work and we will not further comment on this additional boundary term in this paper.

%%%%%%%%%%%%%%%%%%%%%%%%%%%%%%%%%%%%%%%%%%%%%%%%%%%%%%%%%%%%%%%%%%%%%%
\section{Boundary conditions of classical Liouville theory}
\label{app:BC's}
%%%%%%%%%%%%%%%%%%%%%%%%%%%%%%%%%%%%%%%%%%%%%%%%%%%%%%%%%%%%%%%%%%%%%%
In this appendix, we study classical solutions of Liouville theory defined on a disk with all possible boundary conditions.
Since all discussion of this appendix applies to the timelike Liouville theory in the same manner, here we focus only on the spacelike Liouville theory:
	\begin{align}
		S \, = \, \int d^2 x \sqrt{g} \, \Big[ g^{\m\n} \pa_\m \phi \pa_\n \phi \, + \, \m \, e^{2b\p} \Big] \, .
	\end{align}
We use the coordinates (\ref{angular coord}) and in the following we consider static solutions. (Here by static, we mean $\pa_\th \phi =0$.)
Variation of the action on this coordinates can be written explicitly as
	\begin{align}
		\d S \, = \, 2 \int_{r=1}d\th \, (\pa_r \phi) \, \d \phi \, + \, 2 \int dr d\th \, r \, \left[ - \frac{1}{r} \pa_r ( r \pa_r \phi) \, + \, b\m \, e^{2b\p} \right] \d \phi \, .
	\label{delta S}
	\end{align}

In order for the boundary term to vanish, we can take either boundary condition
	\begin{align}
		&{\rm Dirichlet}: \qquad \phi(r=1) \, = \, \phi_0 \quad (= \, {\rm constant}) \\[2pt]
		&{\rm Neumann}: \qquad \pa_r\phi \big|_{r=1} \, = \, 0 \, .
	\end{align}
If we include an additional boundary action
	\begin{align}
		S_{\rm bdy} \, = \, \int_{r=1} d\th \, 4\pi \m_B \, e^{b \phi} \, ,
	\end{align}
we can have the FZZT brane (modified Neumann) boundary condition \cite{Fateev:2000ik,Teschner:2000md}:
	\begin{align}
		{\rm FZZT}: \qquad \pa_r\phi \, + \, 2\pi \m_B b \, e^{b \phi} \, = \, 0 \, , \qquad ({\rm at}\ r=1)
	\end{align}
The ZZ brane boundary condition \cite{Zamolodchikov:2001ah} is a special case of the Dirichlet boundary condition with $\phi_0=\infty$.

The equation of motion can be read off from the second term of (\ref{delta S}) as
	\begin{align}
		 - \frac{1}{r} \pa_r ( r \pa_r \phi) \, + \, b\m \, e^{2b\p} \, = \, 0 \, ,
	\end{align}
and the most general solution is given by
	\begin{align}
		 e^{2b \phi(r)} \, = \, \frac{1}{\m b^2} \frac{c_1^2}{r^2 \sinh[c_1(c_2 - \log r)]^2} \, , 
	\label{general solution}
	\end{align}
where $c_1$ and $c_2$ are integration constants.
We first require a regularity condition at the center of the disk ($r\to 0$).
Since
	\begin{align}
		 e^{2b \phi(0)} \, \sim \, \frac{1}{\m b^2} \frac{4c_1^2}{r^{2-2c_1}} \, , 
	\end{align}
the regularity condition fixes $c_1 =1$.
Now the solution (\ref{general solution}) becomes
	\begin{align}
		 e^{2b \phi(r)} \, = \, \frac{1}{\m b^2} \frac{4c^2}{(1-c^2 r^2)^2} \, , 
    \label{regular solution}
	\end{align}
where we set $c=e^{-c_2}$.

The Dirichlet boundary condition requires
	\begin{align}
		 e^{2b \phi(1)} \, = \, \frac{1}{\m b^2} \frac{4c^2}{(1-c^2)^2} \, = \, e^{2b \phi_0} \, . 
	\end{align}
Therefore, if we set $c^2 =1-p$ (with $\m b^2=1$), we find the solution (\ref{solpc}).
The ZZ brane boundary condition is a special case of this result with $c=1$, which gives 
	\begin{align}
		 e^{2b \phi(r)} \, = \, \frac{1}{\m b^2} \frac{4}{(1- r^2)^2} \, .
	\end{align}
The FZZT brane boundary condition fixes the remaining constant as 
	\begin{align}
		c \, = \, - \frac{2\pi \m_B b^2}{\sqrt{\m b^2}} \, .
	\end{align}
Therefore, the standard Neumann boundary condition ($\m_B=0$) leads to the trivial solution ($c=0$).

%%%%%%%%%%%%%%%%%%%%%%%%%%%%%%%%%%%%%%%%%%%%%%%%%%%%%%%%%%%%%%%%%%%%%%
\section{On-shell actions of Liouville theory}
\label{app:on-shell action}
%%%%%%%%%%%%%%%%%%%%%%%%%%%%%%%%%%%%%%%%%%%%%%%%%%%%%%%%%%%%%%%%%%%%%%
In this appendix, we present some detail of the on-shell actions for the disk geometry discussed in section~\ref{sec:FZZT brane}
and for the punctured disk geometry discussed in section~\ref{sec:annulus amplitude}.

%%%%%%%%%%%%%%%%%%%%%%%%%%%%%%%%%%%%%%%%%%%%%%%%%%%%%%%%%%%%%%%%%%%%%%
\subsection{Disk geometry}
%%%%%%%%%%%%%%%%%%%%%%%%%%%%%%%%%%%%%%%%%%%%%%%%%%%%%%%%%%%%%%%%%%%%%%
The on-shell action for $\phi$ reads
\ba
&& I^{bulk}_\phi=\int d\theta\left[\left(\frac{2}{b^2\delta}+\frac{2}{b^2}(\log 2-1)+\frac{2}{b^2}
\log\delta   - \frac{\delta}{b^2} \right)
+\eta\left(\frac{1}{b^2\delta}-\frac{1}{b^2}+\frac{\delta}{2b^2}\right)\right],  \no
&& I^{bdy}_\phi=\frac{2}{b}\int \phi_0 d\theta=\int d\theta \left[-\frac{2}{b^2}\log\delta 
+\frac{\eta}{b^2}\right],\no
&& I^{ct}_{\phi}=-\frac{2}{b^2}\int d\theta e^{b\phi_0}=-\frac{2}{b^2}
\int d\theta \left[\frac{1}{\delta}+\frac{\eta}{2\delta}\right],
\ea
which give the total action
\ba
I^{bulk}_\phi+I^{bdy}_\phi+ I^{ct}_{\phi} &=& \int d\theta \left[\frac{\delta}{2b^2}\eta+\frac{2}{b^2}(\log 2-1)   - \frac{\delta}{b^2} \right] \no
&=& \int d\theta \left[-A+\frac{2}{b^2}(\log 2 -1) {  - \frac{\delta}{b^2}} \right].
\ea

On the other hand, for the field $\chi$ we find 
\ba
&& I^{bulk}_\chi=\int d\theta\left[\left(-\frac{2}{b^2\delta}-\frac{2}{b^2}(\log 2-1)-\frac{2}{b^2}
\log\delta   + \frac{\delta}{b^2} \right)
+\eta\left(\frac{1}{b^2\delta}-\frac{1}{b^2}+\frac{\delta}{2b^2}\right)\right],  \no
&& I^{bdy}_\chi=\frac{2}{b}\int \chi_0 d\theta=\int d\theta \left[\frac{2}{b^2}\log\delta 
+\frac{\eta}{b^2}\right],\no
&& I^{ct}_{\chi}=-\frac{2}{b^2}\int d\theta e^{-b\chi_0}=-\frac{2}{b^2}
\int d\theta \left[-\frac{1}{\delta}+\frac{\eta}{2\delta}\right],
\ea
which give the total action
\ba
I^{bulk}_\chi+I^{bdy}_\chi+ I^{ct}_{\chi} &=& \int d\theta \left[\frac{\delta}{2b^2}\eta-\frac{2}{b^2}(\log 2-1)  + \frac{\delta}{b^2} \right] \no
&=& \int d\theta \left[-A-\frac{2}{b^2}(\log 2 -1) + \frac{\delta}{b^2} \right].
\ea

%%%%%%%%%%%%%%%%%%%%%%%%%%%%%%%%%%%%%%%%%%%%%%%%%%%%%%%%%%%%%%%%%%%%%%
\subsection{Punctured disk geometry}
%%%%%%%%%%%%%%%%%%%%%%%%%%%%%%%%%%%%%%%%%%%%%%%%%%%%%%%%%%%%%%%%%%%%%%

As we consider bulk one-point function, we include the contribution from the vertex operator into the bulk actions as
	\begin{align}
		I^{bulk}_\phi \, &= \, \int d^2x \sqrt{g} \, \Big[(\de_a\phi)^2 + \mu e^{2b\phi} \Big] \, - \, \a \int d\theta \, \phi(\e) \, , \nn\\
		I^{bulk}_\chi \, &= \, \int d^2x \sqrt{g} \, \Big[-(\de_a\chi)^2 - \mu e^{-2b\chi} \Big] \, - \, \a \int d\theta \chi(\e) \, .
	\end{align}
Since the bulk fields $\phi$, $\chi$ are diverging at $r=0$, we regularize this by integrating $\e\le r \le 1$ and inserting the vertex operators at $r=\e$.
Then, we simply eliminate all diverging contribution in $\e \to 0$ limit.
Now the on-shell action for $\phi$ reads
	\begin{align}
		& I^{bulk}_\phi \, = \, \frac{1}{b^2} \int d\theta \left[\left(\frac{2}{\delta} - \n(2+\n \d) + 2\n \log2\n + 2\log\d \right) \, + \, \eta\left(\frac{1}{\d}-1+\frac{\n^2\delta}{2}\right) \right] \, , \no
		& I^{bdy}_\phi \, = \, \frac{2}{b}\int \phi_0 d\theta \, = \, \frac{1}{b^2} \int d\theta \Big[-2\log\delta + \eta \Big] \, , \no
		& I^{ct}_{\phi} \, = \, - \, \frac{2}{b^2}\int d\theta e^{b\phi_0} \, = \, -\frac{1}{b^2} \int d\theta \left[\frac{2}{\delta}+\frac{\eta}{\delta} \right],
	\end{align}
which give the total action
	\begin{align}
		I^{bulk}_\phi+I^{bdy}_\phi+ I^{ct}_{\phi} \, &= \, \frac{1}{b^2} \int d\theta \left[\frac{\n^2\delta\h}{2}+2\n\log2\n - \n(2+\n\d) \right] \, .
	\end{align}
On the other hand, for the field $\chi$ we find 
	\begin{align}
		& I^{bulk}_\chi \, = \, \frac{1}{b^2} \int d\theta\left[\left(-\frac{2}{\delta} + \n(2+\n \d) - 2\n \log2\n - 2\log\d \right) \, + \, \eta\left(\frac{1}{\d}-1+\frac{\n^2\delta}{2}\right) \right] \, ,  \no
		& I^{bdy}_\chi \, = \, \frac{2}{b}\int \chi_0 d\theta \, = \, \frac{1}{b^2} \int d\theta \Big[2\log\delta + \eta \Big] \, ,\no
		& I^{ct}_{\chi} \, = \, -\frac{2}{b^2}\int d\theta e^{-b\chi_0} \, = \, -\frac{1}{b^2} \int d\theta \left[-\frac{2}{\delta}+\frac{\eta}{\delta} \right],
	\end{align}
which give the total action
	\begin{align}
		I^{bulk}_\chi+I^{bdy}_\chi+ I^{ct}_{\chi} \, &= \, \frac{1}{b^2} \int d\theta \left[\frac{\n^2\delta\h}{2}-2\n\log2\n + \n(2+\n\d) \right] \, .
	\end{align}
Therefore, the total of the spacelike and timelike contributions is 
	\begin{align}
		I_\phi \, + \, I_\chi \, = \, -2\n^2 A \int d\theta \, = \, -4\pi \n^2A \, . 
	\end{align}

%%%%%%%%%%%%%%%%%%%%%%%%%%%%%%%%%%%%%%%%%%%%%%%%%%%%%%%%%%%%%%%
%%%%%%%%%%%%%%%%%%%%%%%%%%%%%%%%%%%%%%%%%%%%%%%%%%%%%%%%%%%%%%%

%----- Bibliography ----------------------
\bibliographystyle{JHEP}
\bibliography{Refs} 

\addcontentsline{toc}{section}{References}

%---------------------------------------------
%---------------------------------------------

\end{document}